\documentclass[journal]{IEEEtran}

\usepackage{cite}
\usepackage{enumerate}
\usepackage[table]{xcolor}
\usepackage[pdftex]{graphicx}
\graphicspath{{../figures/}, {../photo/}}
\usepackage[cmex10]{amsmath}
\usepackage{amsfonts}
\usepackage{multirow}
\usepackage[caption=false,font=footnotesize]{subfig}
\usepackage{stfloats}
\usepackage{url}
\usepackage{algorithm}
\usepackage{algorithmic}
\usepackage{siunitx}
\usepackage{mathrsfs}
\usepackage{array}
\usepackage{verbatim}

\usepackage{booktabs, cellspace}
\let\oldvarphi\varphi
\renewcommand*{\varphi}{\boldsymbol{\oldvarphi}}
\let\oldpsi\psi
\renewcommand*{\psi}{\boldsymbol{\oldpsi}}
\let\oldPhi\Phi
\renewcommand*{\Phi}{\boldsymbol{\oldPhi}}
\let\oldphi\phi
\renewcommand*{\phi}{\boldsymbol{\oldphi}}
\let\oldPsi\Psi
\renewcommand*{\Psi}{\boldsymbol{\oldPsi}}
\let\oldvarTheta\varTheta
\renewcommand*{\varTheta}{\boldsymbol{\oldvarTheta}}
\let\oldalpha\alpha
\renewcommand*{\alpha}{\boldsymbol{\oldalpha}}
\let\oldbeta\beta
\renewcommand*{\beta}{\boldsymbol{\oldbeta}}

\newcommand*{\A}{\boldsymbol{A}}

\newcommand*{\R}{\mathbb{R}}
\newcommand*{\bP}{\boldsymbol{P}}

\newcommand*{\x}{\mathbf{x}}
\newcommand*{\y}{\mathbf{y}}

\newcommand*{\bs}{\mathbf{s}}

\usepackage{scalerel}
\usepackage{tikz}
\usetikzlibrary{svg.path}

\definecolor{orcidlogocol}{HTML}{A6CE39}
\tikzset{
  orcidlogo/.pic={
    \fill[orcidlogocol] svg{M256,128c0,70.7-57.3,128-128,128C57.3,256,0,198.7,0,128C0,57.3,57.3,0,128,0C198.7,0,256,57.3,256,128z};
    \fill[white] svg{M86.3,186.2H70.9V79.1h15.4v48.4V186.2z}
                 svg{M108.9,79.1h41.6c39.6,0,57,28.3,57,53.6c0,27.5-21.5,53.6-56.8,53.6h-41.8V79.1z M124.3,172.4h24.5c34.9,0,42.9-26.5,42.9-39.7c0-21.5-13.7-39.7-43.7-39.7h-23.7V172.4z}
                 svg{M88.7,56.8c0,5.5-4.5,10.1-10.1,10.1c-5.6,0-10.1-4.6-10.1-10.1c0-5.6,4.5-10.1,10.1-10.1C84.2,46.7,88.7,51.3,88.7,56.8z};
  }
}

\newcommand\orcidicon[1]{\href{https://orcid.org/#1}{\mbox{\scalerel*{
\begin{tikzpicture}[yscale=-1,transform shape]
\pic{orcidlogo};
\end{tikzpicture}
}{|}}}}

\usepackage{hyperref} 

\begin{document}

\title{Privacy Assured Recovery of Compressively Sensed ECG signals}

 
   \author{Hadi Zanddizari \orcidicon{0000-0002-2465-9374},~\IEEEmembership{Student Member,~IEEE,}
      Sreeraman Rajan \orcidicon{0000-0003-0153-6723},~\IEEEmembership{Senior Member,~IEEE,}	
      Hassan Rabah \orcidicon{xxx0-xxxx-xxxx-xxxx},~\IEEEmembership{Senior Member,~IEEE,} 	
      Houman Zarrabi \orcidicon{0000-0002-5890-7321},~\IEEEmembership{Senior Member,~IEEE,}

  \thanks{Hadi Zanddizari is with the Department of Electrical Engineering, University of South Florida, USA (e-mail: hadiz@mail.usf.edu).}
  \thanks{}   
 
 \thanks{\textcopyright 2019 IEEE. Personal use of this material is permitted.  Permission from IEEE must be obtained for all other uses, in any current or future media, including reprinting/republishing this material for advertising or promotional purposes, creating new collective works, for resale or redistribution to servers or lists, or reuse of any copyrighted component of this work in other works.}}

\markboth{}{}
\maketitle

\begin{abstract}

	Cloud computing for storing data and running complex algorithms have been steadily increasing. As connected IoT devices such as wearable ECG recorders generally have less storage and computational capacity, acquired signals get sent to a remote center for storage and possible analysis on demand. Recently, compressive sensing (CS) has been used as secure, energy-efficient method of signal sampling in such recorders. In this paper, we propose a secure procedure to outsource the total recovery of CS measurement to the cloud and introduce a privacy-assured signal recovery technique in the cloud. We present a fast, and lightweight encryption for secure CS recovery outsourcing that can be used in wearable devices, such as ECG Holter monitors. In the proposed technique, instead of full recovery of CS-compressed ECG signal in the cloud, to preserve privacy, an encrypted version of ECG signal is recovered by using a randomly bipolar permuted measurement matrix. The user with a key, decrypts the encrypted ECG from the cloud to obtain the original ECG signal. We demonstrate our proposed method using the ECG signals available in the MITBIH Arrhythmia Database. We also demonstrate the strength of the proposed method against partial exposure of the key. 
\end{abstract}

\section{Introduction}

\IEEEPARstart{I}{n}  biomedical  area,  there  are  equipment  and  devices  that produce  huge  amount  of  data.  As  an  example,  Holter monitor,  a  wearable  device,  is  used  for  continuous monitoring  of  the  electrical  activity  of  the  heart,  namely electrocardiogram  (ECG). Holter monitor is used by cardiac patients for several days to capture events such as cardiac arrhythmia. Even in cases, where physiological signals are recorded only intermittently, say by devices such as Empatica watch that have limited on the device storage memory, the amount of data produced is large that an external storage solution is in order. The ECG data produced by devices that continuously monitor, need to be stored for analysis and tracking improvements in the physiology when medical interventions are undertaken. Compression can be used for efficient use of communication channel bandwidth and storage. 

Recently compressive sensing (CS) has been used as a fast and energy-efficient algorithm for simultaneous sampling and compressing of potentially sparse signals~\cite{01Baraniuk2007, 02Donoho2006}. CS has wide variety of applications in signal processing such as biomedical signal compression, enhancement, and recovery \cite{Enhancement,Khosh}. The applications of CS has also been extended to ECG signal under the assumption that ECG signals are compressible signal ~\cite{03Mamaghanian_2011,04Pant2014,05Polania2015,06Abo_Zahhad_2015,07Dixon2012,08Mishra2012,09Zanddizari2018,10Polania2011}. In~\cite{03Mamaghanian_2011}, CS-based compression was shown to present the best overall energy efficiency due to its lower complexity and also reduced CPU execution time. Since compression phase in CS is simple, fast, and energy efficient, CS has been chosen for compression in many sensing applications. However, recovery phase which is non-linear and complex in terms of computations demands use of processors that have speed, large on-board memory and computational capability.
Currently wearable devices do not have such capabilities and storage and recovery need to be outsourced. In addition, clinics and hospitals, usually generate enormous amount of data, thereby requiring a place to store the data. Cloud environments generally provide “unlimited” resources and facilities. Hence, cloud can be used for storing CS-based compressed ECG signal, and based on user request ECG signal can be recovered. However, the cloud as a third party between real user (patient) and clinic should not be permitted to have access to the recovered ECG signal which will be referred to as the plaintext in this paper. 

There are numerous works that have proposed different procedures for secure CS recovery outsourcing~\cite{18Wang_2013,19Xue_2017,42Zanddizari2015}. 
There are some research papers that specifically  focus on methods for assuring the secrecy of ECG data in communication~\cite{20Liu2018,21Fira2015,22Sufi2008}. Recently, T.Y. Liu et al proposed a new encryption-then-compression (ETC) method for the ECG signal~\cite{20Liu2018}. In this work, authors encrypt ECG signal with a symmetric key. Their proposed key is a square orthogonal random matrix that changes after sending every encrypted ECG signal (ciphertext). Hence, their method can be classified as a one-time-pad cryptosystem. For compression, they apply a transformation based algorithm in which singular value decomposition (SVD) technique is used~\cite{23Wei2001}. However, it just focuses on compression and does not consider the security of the data. In~\cite{20Liu2018}, a modification has been made on SVD technique to provide secrecy as well. SVD-based methods bring higher compression ratio (CR). They are computationally intensive and may introduce delay in the system. In contrast, at expense of having lower compression ratio, CS-based methods are linear, faster, and energy efficient~\cite{24Karakus2013}.

This paper presents a novel fast, and light-weight encryption for secure CS recovery outsourcing that can be used in resource constrained devices, such as wearable ECG recorders. 
The paper is arranged in the following manner.  In next section, CS is introduced and presented as a cryptosystem, and followed by the background research in secure CS recovery outsourcing. Section~\ref{sec_Proposed_method} contains the proposed method followed by security analysis and experimental results to verify the secrecy of the method in section ~\ref{sec_results}.  Finally Section ~\ref{sec_conclusion} concludes the paper.

 \section{Background}
 \label{sec_Background}
 \subsection{Compressive sensing}
 \label{sebsec_cs}
 
 Compressive  sensing  (CS)  is  a  sampling  technique  for efficiently  sampling  a signal  by  solving under-determined linear  systems~\cite{01Baraniuk2007,02Donoho2006}. It  takes  advantage  of  the  signal's sparsity, and the signal can be effectively represented by fewer number of measurements than the Nyquist rate. For instance, given an ECG signal $\x \in \R^N$ and an orthogonal basis $\Psi \in \R^{N \times N}$, then one can map the ECG signal to sparse domain via, $\x=\Psi s$, where $\bs \in \R^N$ is sparse vector with $k$ ($ k << N$) nonzero entries. 
 In other words,  $\bs$  is a sparse representation of  $\x$  under the chosen predefined dictionary. Compression phase in CS provides the measurement vector through a linear operation as given below: 
 
 \begin{equation}
 \y = \Phi \x =  \Phi \Psi \bs \label{equ_11}
 \end{equation}
 where, $\mathbf{y} \in \R^{M}$ is the measurement vector and $\Phi \in \R^{M \times N}$ is the measurement matrix. For simplicity, let  $\mathbf{A} = \Phi \Psi$. $\mathbf{A} \in \R^{M \times N} $ is a rectangular matrix, sometimes referred to as “total” dictionary in the CS literature. For exact and stable recovery of sparse signal,  \textit{restricted isometry property} (RIP) is a sufficient condition \cite{25Candes2008}.
RIP is satisfied if there exists a restricted isometry constant (RIC) $\delta_K$ , $0 < \delta_K < 1$ such that
 
 \begin{equation}
 \label{eq_delta}
 (1 - \delta_K) \|  \mathbf{s}  \|_2^2 \le \| \mathbf{A}  \mathbf{s}  \|_2^2 \le (1 + \delta_K) \|  \mathbf{s} \|_2^2
 \end{equation}
 where $\delta_K$ denotes  isometry  constant  of  a  matrix  $\bf{A}$,  and  its  value  belongs  to  a  set  of  real  numbers  between  zero  and  one. But, checking the RIP condition of a matrix or calculating the value  of  its  isometry  constant  is  difficult to verify.
 ~~Hence, conditions that lead to RIP were proposed~\cite{15Needell2008,26Tropp_2007}. Another condition, which is easier to verify in practice, is the requirement that measurement matrix $\Phi$ must be incoherent with the sparsity basis $\Psi$. Mutual coherence $\mu$ between $\Phi$ and $\Psi$ is defined as follow:
 
 \begin{equation}
 \label{eq_muphipsy}
 \mu (\Phi, \Psi) = \sqrt{N} \max_{i, j} \frac{| \left\langle  \phi_i, \psi_j \right\rangle  |}{ \| \phi_i \|_2 \| \psi_j \|_2}
 \end{equation}
 where $\phi_{i \in \left\lbrace 1, \hdots , M \right\rbrace }$ and $\psi_{j \in \left\lbrace 1, \hdots , N \right\rbrace }$ respectively represent the row vectors of $\Phi$ and the column vectors of $\Psi$. The coherence measures the maximum correlation between the two matrices. Smaller coherence can lead to better signal reconstruction performance~\cite{50Yan2014}. Since $\mu \in [ 1, \sqrt{N} ]$, the matrices $\Phi$ and $\Psi$ are incoherent if $\mu (\Phi, \Psi)$ is closer to one, which corresponds to the lower bound of $\mu$.

 A step called the \textit{recovery process} reconstructs the input signal $\x$ from the measurement vector $\y$ by solving the equation (\ref{equ_11}).
 Since $\A$ is a rectangular matrix ($M < N$), the problem formulated in equation (\ref{equ_11}) is ill-posed and has infinite solutions.
 However, based on the knowledge that $\x$  has a sparse representation with respect to a basis $\Psi$, the recovery process can be performed in two steps~\cite{25Candes2008}. The first step finds the sparse vector $\tilde{\bf{s}}$ by solving the following equation:
 
 \begin{equation}
 \label{equ_2}
 \min\limits_{\tilde{\bs}}\|\tilde{\bs}\|_0 \textrm{ such that } \A \tilde{\bs} = \y. 
 \end{equation}
Once the vector $\tilde{\bs}$ has been obtained, the second step reconstructs the original signal as follows:
 
 \begin{equation}
 \tilde{\x} = \Psi \tilde{\bf{s}}.
 \end{equation} 

Various methods have been proposed to find an appropriate solution to equation~(\ref{equ_2}) leading to numerous recovery algorithms such as Basic Pursuit \cite{11Chen2001,12Dai2009}, StOMP \cite{13Donoho_2012}, OMP~\cite{14Patia}, CoSAMP~\cite{15Needell2008}, Belief Propagation~\cite{16Baron2010} and SL0~\cite{17Mohimani2009}. 
 
 \subsection{CS as cryptosystem}
 \label{subsec_cs_crypto}

 From a different viewpoint, CS can be assumed as a cryptosystem \cite{one_Time}. Since CS can map every sparse signal from $N$ dimensional space to $M$ dimensional space, where $M \ll N $, numerous researchers have considered CS as a strong cryptosystem \cite{CS_crypto,28Cambareri2015,29Yuan2016,30Yang2015,31Bianchi2016,32Mangia2018,33Zhang2018,34Barcelo-Llado2012,35Chu2015,36Zhang2011,37Lu2013,38Mayiami_2013}.  In \cite{38Mayiami_2013}, it was proved that under certain conditions, CS can even meet the perfect secrecy as defined by Shannon. 
 
 In \cite{LFSR1,LFSR2}, Linear Feedback Shift Registers have been used to generate CS measurement matrix as a key. In \cite{low_complexity}, a low-complexity approach for Privacy-Preserving Compressive Analysis based on subspace-based representation has been proposed to preserve privacy from an information theoretic perspective.

 One important issue in almost every previous work is that the CS-recovery is assumed to be done by the real user. In other words, by having both the key (measurement matrix) and the ciphertext (measurement vector), a real user would be able to reconstruct the plaintext (initial signal).  Aforementioned  works suppose that decryption can be done at the user end. But, in many contexts, devices at the user end do not have enough computational resources; hence such complex decryption is not feasible. A powerful remote server or cloud can be used for doing recovery process of CS problem. However, the third party would have access to the plaintext after the recovery. In addition, while transmitting the recovered signal from the third party back to the user, the data need to undergo encryption again. Privacy preserving outsourcing techniques can be used to overcome these concerns when the recovered signal contains private information of an individual. For example, ECG signal contains information that enable unique identification of an individual. Recently, there have been increased interest on ECG for biometric recognition \cite{39Fratini2015}. Temporal features, amplitude features and morphological features of an ECG signal have been used for ECG-based biometric. As ECG signal contains biometrics of the person, privacy is exposed when ECG signals are fully recovered in the cloud. Hence, a solution that avoids complete recovery to preserve privacy is in order.  Cloud environments can be a good option to store the compressed data. But on-demand through CS-recovery, the plaintext should not be exposed in the cloud. This paper provides a privacy assured CS outsourced recovery. Researchers have proposed different methods to shift away the recovery phase of CS in a secure manner\cite{18Wang_2013,19Xue_2017,42Zanddizari2015}. In this paper, we propose a fast and lightweight privacy-preserved CS recovery approach for ECG signal.

 \subsection{Privacy-preserved CS-Recovery Outsourcing}
 \label{subsec_secure_out}
 
 In privacy-preserved CS-recovery outsourcing, there are three levels of data to be considered: the cipher, the intermediate cipher, and the plaintext. The cipher is the measurement vector which is compressed or encrypted. 
 Since cloud should not obtain the plaintext, recovery in the cloud yields an encrypted signal. This encrypted signal is called intermediate cipher. Intermediate cipher is sent to the real user, and real user (let us say Alice) with a private key can decrypt this intermediate cipher to obtain the plaintext. Plaintext is the original raw signal that got compressed initially.
 
 Recently, "\textit{Outsourced Image Recovery Service (OIRS)}" was proposed by Cong Wen et al \cite{18Wang_2013} where a technique to securely shift away the recovery of CS in cloud environment is presented. However, the proposed method in \cite{24Karakus2013} requires the cloud to solve linear programing (LP) problem to reconstruct the CS-encrypted image (the cipher). In other words, OIRS requires the cloud to use LP method to convert cipher to an intermediate cipher. But, LP is only one of the CS recovery methods and its order of complexity is $\mathcal{O}({N^3})$. However, there are other efficient and faster algorithms from greedy algorithms such as OMP with $\mathcal{O}({kMN})$, or SL0 with $\mathcal{O}({MN})$ that can be used instead of LP method. In addition, OIRS uses multiple keys for assuring privacy. This leads to heavy computation and consumes large time for processing. Such algorithms may not be appropriate for simultaneous encryption and compression of wearable ECG recorders where we have limited power and computational capability.  "\textit{Kryptein}" is another CS-based encryption scheme for the internet of things (IoT) that has been proposed by Xue et al \cite{19Xue_2017}. In this work, CS has been used as compression and encryption algorithm. The secrecy of proposed cryptosystem mainly revolves around the sparsifying dictionary. However, it limits CS by choosing the adaptive sparsifying dictionaries. In other words, it uses an adaptive dictionary learning to generate sparsifying dictionary and also use this learnt dictionary as part of the key. This learnt sparsifying dictionary along with a perturbation matrix are used for designing their secret key. In \cite{42Zanddizari2015}, a secure reconstruction of image from CS in cloud was introduced.It assumes CS as a compression algorithm, and not as a sampling method. The pre-processing used in this work led to delays in generating compressed and encrypted signal. This method mapped the initial signal to a sparse domain and then put a threshold to force negligible coefficients to be zero. This method, thus, required all the components of sparse vector to be checked for zeroing, which may not be an efficient way of utilizing the limited computational capability and the power of weak devices such as ECG wearable recorders. 
 
 In this paper a simple, light-weight encryption is applied to map the initial sparse signal to another sparse signal. Considering the fact that sparsity is a required condition for CS, we are limited in options as we cannot violate this condition. In order to maintain sparsity and still achieve light weight encryption, two keys are used:  a random square matrix and  a random bipolar permutation matrix. The former encrypts the measurement matrix uniquely for each wearable recorder, and the latter encrypts signal after reconstruction in the cloud for secure transmission back to the user.
 
 %

\section{Proposed method}
\label{sec_Proposed_method}

Consider a common scenario where an ECG sensor sends $\mathbf{y} = \Phi \mathbf{x}  =  \Phi \Psi \mathbf{s} $  to cloud environment for storage. For simplicity, let  us  suppose $\mathbf{A} = \Phi \Psi $.  On  demand  for  recovery,  cloud  can reconstruct  the  sparse  signal  $\mathbf{s}$  if  the  cloud  is supplied with both  $\mathbf{y}$ and  $\mathbf{A}$ . Cloud can choose any CS recovery algorithm to solve the following  $\ell_1$ \textit{minimization}  problem:

\begin{equation}
\min\limits_{\mathbf{s}}\|\mathbf{s}\|_1 \textrm{ s. t. } \mathbf{A}  \mathbf{s} = \mathbf{y}. \label{equ_3}
\end{equation}

Once $\bs$ is obtained, the initial ECG signal can be generated using $\Psi$, i.e; $\x=\Psi \textbf{s}$. In order to securely shift away the full CS-recovery task, the use of two keys is proposed. The first key is used to encrypt $\Phi$. Because,  measurement  matrix  is  a  specific  information  of every  CS-based sensing device, it should not be shared  with the  third  party.  In  addition,  besides  random  class  of measurement matrices that preserve RIP condition, there is also a deterministic  approach  to  generate  a  measurement  matrix \cite{DBBD,44Applebaum2009,45Calderbank2010}. Since such matrices have defined structures, if we encode  these  structures,  then  we  may increase  the  secrecy  of cryptosystem.  To  do  so,  we  use  a  random  measurement  matrix  $\mathbf{Q}_{M \times M}$   to  encrypt  initial  measurement  matrix.  Then, instead of sending $\mathbf{A}$ to the cloud, $\hat{\mathbf{A}} =\mathbf{Q}(\mathbf{\Phi} \mathbf{\Psi})=\mathbf{Q} \mathbf{A}$ will be sent. If we multiply $\mathbf{Q}$, the recovery relation is changed as follows:

\begin{equation}
\min\limits_{\bs}\|\bs\|_1 ~~~\textrm{ s. t. } \hat{\mathbf{A}} \bs = \mathbf{Q} \y =\hat{\y}. \label{equ_4}
\end{equation}

When $\hat{\mathbf{A}}$ and $\hat{\y}$ are provided to the cloud,  the cloud can reconstruct the sparse vector  $\bs$. This level of encryption  just  hides  the  measurement  matrix but the secrecy  of  reconstructed  signal  is still  not  preserved.  To  further maintain  secrecy, a second key is used in the following manner. We multiply the encrypted measurement matrix with a random bipolar permutation matrix $\mathbf{P}$, an invertible matrix that contains either ``$\alpha$ or $-\alpha$" in each row and column at random positions,- where $\alpha$ is a random scalar number. For example, a 5x5 $\mathbf{P}$ may be as follows

\begin{equation}\label{eq_P}
\mathbf{P}_{5 \times 5}=
\begin{bmatrix}
0 & -\alpha & 0 & 0 & 0 \\ 
+\alpha & 0 & 0 & 0 & 0 \\ 
0 & 0 & +\alpha & 0 & 0 \\ 
0 & 0 & 0 & 0 & -\alpha \\ 
0 & 0 & 0 & +\alpha & 0
\end{bmatrix} 
\end{equation}

By multiplying $\hat{\mathbf{A}}$  with $\mathbf{P}$, a resultant new matrix $\mathbf{A}^*= \hat{\mathbf{A}} \mathbf{P}$ results. The effect of $\mathbf{P}$ is to map the reconstructed sparse signal into a random permuted sparse signal and to randomly change the sign of the sparse components. After this multiplication, the recovery in cloud becomes:

\begin{equation}
\min\limits_{\mathbf{P}^{-1}\bs}\|\mathbf{P}^{-1}\bf{s}\|_1 \textrm{ s. t. } (\hat{\mathbf{A}} \bP)(\mathbf{P}^{-1}\bs) = \mathbf{A}^{*}(\mathbf{P}^{-1}\bs)=\hat{\y} \label{equ_14}
\end{equation}

Sending the  $\mathbf{A}^*$   and  $\hat{\y}$  to the  cloud,  the cloud  would be able to  recover  the  intermediate cipher,  $\mathbf{P}^{-1}\bs$. Since the inverse of a permutation matrix is also a permutation matrix, the recovered signal from cloud is still sparse.  Note  that  $\mathbf{P}$   changes  the  position  of  components  and does not change the order of sparsity. Therefore, we can guarantee that the  sparsity  of  signal  is  preserved.  Sparsity  is  a  required condition for CS recovery, and without it, recovery cannot be done accurately. In words, in our proposed method, the original sparse vector is now mapped into another sparse vector; this mapping is done in the sparse domain and not in the domain in which signal is acquired. In the proposed privacy-assured  recovery,  the cloud  after  recovery  yields  $\mathbf{P}^{-1}\bs$,  which  is  a mapped sparse vector or encrypted sparse vector. Cloud  may then send the encrypted sparse vector,  $\mathbf{P}^{-1}\bs$,   to  the real  user,  and  the  user  would  be  able  to reconstruct initial signal by using corresponding key,  $\mathbf{P}$  as follows: 

\begin{equation}
\mathbf{P} *\mathbf{P}^{-1}\bs = \bs ;~~~~~ \Psi \bs = \x \label{equ_15}
\end{equation}
Note that $\mathbf{P} = \alpha\mathbf{P'}$ where $\mathbf{P'}$ is an orthonormal matrix. Multiplication of an orthonormal matrix with a measurement matrix  does  not  affect  the  RIP  condition. 
Multiplying  $\hat{\mathbf{A}}$  by $\mathbf{P}$  would still preserve the RIP inequality:

\begin{equation}
(1 - \delta_K) \| \mathbf{P} \bs \|_2^2 \le \|  \hat{\mathbf{A}} \mathbf{P} \bs \|_2^2 \le (1 + \delta_K) \| \mathbf{P} \bs \|_2^2.
\end{equation}
As $\mathbf{P} = \alpha\mathbf{P'}$, the above equation can be rewritten as follows:

\begin{equation}
(1 - \delta_K) \| \mathbf{P'} \bs \|_2^2 \le \|  \hat{\mathbf{A}} \mathbf{P'} \bs \|_2^2 \le (1 + \delta_K) \| \mathbf{P'} \bs \|_2^2
\label{eq_orth}
\end{equation}
Note that  $ \| \bs \|_2^2 = \| \mathbf{P'} \bs \|_2^2 $, that is, $\mathbf{P'}$ does not change the norm of a sparse vector. In this case, left and right sides of inequality shown in equation \ref{eq_orth} would be same as without encryption mode which means, $\mathbf{P}$ does not affect the RIP condition.

\section{Results}
\label{sec_results}

In order to assess the quality of reconstruction, appropriate metrics need to be considered. There are a few metrics proposed in the literature to  measure  the  quality  of  reconstructed  signal. Three such metrics that are commonly used for assessing the quality of recovered ECG signals are \textit{percentage root-mean-square difference} (PRD), the normalized version of PRD namely PRDN, and signal to noise ratio (SNR),   

\begin{equation}
PRD [\%] = 100 \sqrt{\frac{ \sum^{N-1}_{n=0}(x(n) - \tilde{x}(n))^2}{\sum^{N-1}_{n=0} x^2(n)},}
\end{equation}

\begin{equation}
PRDN [\%] = 100 \sqrt{\frac{ \sum^{N-1}_{n=0}(x(n) - \tilde{x}(n))^2}{\sum^{N-1}_{n=0}(x(n) - \bar{x}(n))^2},}
\end{equation}

\begin{equation}
SNR [\si{\decibel}] = -20 \log_{10}\left(  \frac{PRD}{100} \right), 
\end{equation}
where  $x(n)$ is the original signal, $\tilde{x}(n)$ is the recovered signal, $\bar{x}(n)$ is the mean  of original ECG signal (uncompressed), and $N$ denotes the length of ECG signal. In \cite{27Zigei2000}, Zigel et al. established a link between the PRD and the diagnostic distortion. In \cite{27Zigei2000}, different values of PRD for the reconstructed ECG signals were considered and a qualitative assessment as perceived by the specialist was given. Table \ref{tab_2} shows the classified quality and corresponding PRD and SNR. 

\begin{table}[hbtp]
	\setlength\abovecaptionskip{-0.1\baselineskip}
	\renewcommand{\arraystretch}{1.5}
	\label{tab_2}
	\centering
	\caption{Assessment of Quality of Reconstruction\cite{27Zigei2000}}
\begin{tabular}{p{24mm}|p{24mm}|p{24mm}}
    
	\hline 
	\textbf{PRD}  & \textbf{SNR}   & \textbf{Quality} \\ 
	\hline\hline
	$0 < PRD < 2 \% $& $SNR > 33~dB$ & "Very Good"  \\ 
	\hline 
	$2\% < PRD <9\%$ & $20~dB < SNR < 33~dB$ & "Good"  \\ 
	\hline 
	$PRD \geq 9\%$ & $SNR \leq 20~dB$ &  "undetermined" \\ 
	\hline 
\end{tabular} 
\end{table}

\subsection{Analysis of attacks on intermediate cipher}
Cloud should have a pair of ($\A^*, \hat{\y} $) to conduct the recovery process. Two  scenarios,  one  obtaining $\bP$ based  on  $\A^*$ and  the other obtaining $\bP$  based   on   the recovered   signal   or intermediate  ciphertext are considered.  In the first scenario, it is  statistically  impossible  to  separate  $\bP$ from $\A^*$.  To prove this, we consider a simpler condition where there is no first key. The measurement   matrix  assumed to be an  i.i.d. Gaussian  matrix  with $\mu_{ij}=0$ and $\sigma_{ij}=1/M$,  where  $\mu_{ij}$ and $\sigma_{ij}$ are  the  mean  and  standard  deviation  of  the  i.i.d.  Gaussian matrix  entries,  respectively.  As  the  distribution of the linear  combination  of multiple   independent   random   variables   having   a   normal distribution is also a normal distribution, $\A^*= \Phi\Psi\bP $ is also a Gaussian matrix. The entries of  $\A^*$ and the mean and variance of its entries are obtained as follows:

\begin{equation}
\mathbf{A}^*_{ij}=(\Phi \Psi \mathbf{P})_{ij}=\sum^N_{k=1}\Phi_{ik}(\Psi\mathbf{P})_{kj}
\end{equation}

\begin{equation}
\mathop{\mathbb{E}}(\mathbf{A}^*_{ij})=\sum^N_{k=1}\mu_{ij}(\Psi\mathbf{P})_{kj}=0
\end{equation}

\begin{equation}
\begin{split}
Var(\mathbf{A}^*_{ij})&=\sum^N_{k=1}\sigma^2_{ij}(\Psi\mathbf{P})_{kj}^2\\
&=(\alpha/M)^2\sum^N_{k=1}(\Psi\mathbf{P})_{kj}^2\\
&=(\alpha/M)^2\sum^N_{k=1}(\Psi)_{kj}^2\\
&=\dfrac{\alpha^2\beta}{M^2}
\end{split}
\end{equation}
where subscript $ij$ refers to the element of ith row and jth column of the matrix. $\beta$ is  the  Euclidean  norm  of  the  rows  of  sparsifying dictionary, and  $\mathop{\mathbb{E}}(.)$  and  $Var(.)$  are the mean and variance of random  variable.  Almost  all  sparsifying  dictionaries  are orthonormal,  $\beta=1$ ;  then,  the resultant matrix in the cloud side is a  Gaussian matrix  with  zero  mean  and  variance  $ \alpha^2/M^2$. Also, the covariance of $\mathbf{A}^*$ can be calculated as follow,

\begin{equation}
\begin{split}
Cov(\mathbf{A}^* )&=\mathop{\mathbb{E}}((\Phi\Psi\bP)(\Phi\Psi\mathbf{P})^T)\\
&=\mathop{\mathbb{E}}(\Phi\Psi\mathbf{P}\mathbf{P}^T\Psi^T\Phi^T)\\
&=\mathop{\mathbb{E}}(\Phi(\alpha^2\mathbf{I})\Phi^T)\\
&= \alpha^2\mathop{\mathbb{E}}(\Phi\Phi^T)=\alpha^2Cov(\Phi)\\
&=\dfrac{\alpha^2}{M^2}\mathbf{I}
\end{split}
\end{equation}
where superscript $T$ denotes the matrix transpose, and $\mathbf{I}$ is the identity matrix. Since the entries of $\Phi$ were chosen from an i.i.d Gaussian distribution, the covariance matrix of $\mathbf{A}^*$ is a diagonal matrix which shows its entries are i.i.d as well.   Therefore,  the statistical  distance  of   $\mathbf{A}^*$   in  cloud  and  any  Gaussian  matrix  $\mathcal{N}(0, \alpha/M)$ is zero. In other words, given   $\mathbf{A}^* = \Phi\Psi\mathbf{P}$, cloud cannot  reveal  any  information  about  $\mathbf{P}$,  and  there  is 
  no  statistical difference  between  $\Phi\Psi\mathbf{P}$   and  any  random 
Gaussian  matrix  $\mathcal{N}(0,\alpha/M)$.   

In  the  second  scenario,  a "curious"  cloud  or  an  attacker  tries  to  discover  $\mathbf{P}$   based  on intermediate  ciphertext.  Given  the  intermediate ciphertext, $\mathbf{P}^{-1}\bs$ ,  the  initial  ECG  signal  cannot  be  obtained by $\Psi$ ,  because  $ \x=  \Psi \bs$   and  not   $\Psi\mathbf{P}^{-1}\bs$.  Meanwhile,  an attacker or curious cloud may try to find the bipolar matrix and reconstruct the plaintext or initial uncompressed ECG signal. To do this, attacker should exactly detect the bipolar permutation key. Any change in original key will be completely propagated into the actual time domain values of the signal and corrupt the signal. In other words, as bipolar permutation matrix is applied in sparse domain, the position and sign of elements of sparse vector are changed arbitrarily. After transforming back into the time domain, the recovered signal will be totally different from the original signal. Hence, a small change in permutation matrix can lead to a small change in sparse domain, but a major change in the domain in which signal acquired (generally time domain). To demonstrate the role of bipolar permutation matrix in maintaining the secrecy, recovery was tested with a number of estimated bipolar permutation matrices with different levels of similarity with the original key.  Let the estimated bipolar permutation matrix be  $\mathbf{E}^r$ which $\mathbf{E}$ contains exactly $r\%$ of the columns of the $\mathbf{P}$ and only $(100-r)\%$ of its columns is unclear or unknown for the attacker. Intermediate ciphertext with different estimated permutation matrices (estimated key) were decrypted and the similarity of the estimated key with the actual key, were measured using Frobenius norm. Given a $M \times N$ matrix $\mathbf{A}$, its Frobenius norm is defined as the square root of the sum of the absolute squares of its elements as follow,

\begin{equation}
\| \mathbf{A} \|_F  =  \sqrt{\sum_{i=1}^{M} \sum_{j=1}^{N} |\mathbf{A}_{ij}|^2 }
\end{equation}
where the $\mathbf{A}_{ij}$ is the element of ith row and jth column of $\mathbf{A}$. Accordingly, the Frobenius norm of the difference between the true key and the estimated key can be obtained as follow,

\begin{equation}
\| \mathbf{P} - \mathbf{E}^r \|_F  =  \sqrt{\sum_{i=1}^{M} \sum_{j=1}^{N} |\mathbf{P}_{ij}-\mathbf{E}^r_{ij}|^2 }
\label{frob}
\end{equation}
where the $\mathbf{P}_{ij}$ and $\mathbf{E}^r_{ij}$ are the elements of ith row and jth column of $\mathbf{P}$ and $\mathbf{E}^r$, respectively. Equation \ref{frob} was considered as a metric to show similarity of the estimated key to the actual key. Three estimated matrices were generated by copying 99\%,  98\% and  97\% of the columns of the $\mathbf{P}$ into 3 estimated matrices $\mathbf{E}^{99}$, $\mathbf{E}^{98}$, and $\mathbf{E}^{97}$, respectively. Then, the elements of rest of $1\%$, $2\%$, and $3\%$ columns of these estimated matrices were randomly generated. One ECG signal, record number 101 was selected from the MIT Arrhythmia database \cite{47physiobank}.  First 1000 samples of this ECG signal was selected  
plaintext $\x$, and the orthogonal Daubechies wavelets (db 10) was considered as sparsifying dictionary. Daubechies wavelet (db 10) is the most popular wavelet basis that used in ECG transform-based compression techniques \cite{46Addison2005}. A random bipolar permutation matrix of size $\mathbf{P}_{1000 \times 1000}$ was chosen, and the estimated keys were generated accordingly. The simulation results are available in Table \ref{tab_3} and it shows that a small difference in permutation matrix (or a small dissimilarity) leads to a major difference in decrypted ECG signal. For instance, the $\mathbf{E}^{99}$ contained $99\%$ of the columns of actual key, and just $1\%$ of its columns were chosen randomly. In other words, $990$ columns were the exact replica of the main key, and just $10$ columns were randomly estimated. The simulation results show that these $10$ columns contributed to a totally different decrypted signal from the originally considered plain text ECG signal. 

\begin{table}[!ht]
	\renewcommand{\arraystretch}{1.5}
	\caption{The strength of bipolar permutation key }
	\label{tab_3}
	\centering
	\begin{tabular}{c|cc}
		\hline
		Key 		  & $\| \mathbf{P} - \mathbf{E} \|_F $ & \textbf{PDRN(\%)}\\
		\hline  
		$\mathbf{P}$           & 0 			        & 30		   \\
		$\mathbf{E}^{99}$        & 4.47			    & 249	   \\
		$\mathbf{E}^{98}$        & 6.32			    & 370	   \\
		$\mathbf{E}^{97}$        & 7.73			    & 450	   \\ 
		\hline
	\end{tabular}
\end{table}

Table \ref{tab_3} shows that the permutation key is very sensitive and a small change in its elements can fail to provide exact decryption. Also, consider the scenario that, for instance $\mathbf{E}^{99}$, $99\%$ of its columns are truly estimated. However, in practice such estimation demands heavy computational resources and time. Because, there are $2^N\times N!$ bipolar permutation matrices of size $N$, where $!$ represents the  factorial operation, need to be tried. For the case of $N=1000$, $512$, $256$, or $128$, the number of permutation matrices is a very huge number\footnote{$1000! \cong 4 \times 10^{2567},~128! \cong 3.8 \times 10^{215} $}, and the probability of estimating actual key is negligible. However, considering systematic attack scenario, an unfaithful cloud or eavesdropper may try to employ the order of sparsity ($k$) as a side information and then estimate the initial sparse vector. Although employing the order of sparsity might decrease the search space for attacker, but the attacker still faces a problem of searching for a solution, that is nearly not feasible, that is, for a large $N$ and for a certain given $k$, the search requires non-polynomial ($NP$) time to solve. To estimate the initial sparse vector attacker will have to do $2^k {{N}\choose{k}} $ times exhaustive search. To clarify the complexity of breaking the cipher, if the recovered sparse signal has $1024$ components, and if the sparse vector has at least $64$ nonzero elements (our experiments show more than 64); then it requires $2^{64}{ {1024}\choose{64}} \cong 4.8 \times 10^{102}$  trials for the attacker to guess the plaintext. Moreover, the diagnostic information of an ECG signal is very sensitive, and a bit change in recovered signal can disturb the real information within the signal. The ultimate goal of the proposed method was to provide a simple and secure outsourcing method that is robust to the aforementioned issues. In comparison to one of the strongest outsourced CS-recovery service proposed in \cite{18Wang_2013}, this method of encryption demands less computational resources. In \cite{18Wang_2013}, cloud had to do CS-recovery based on LP method, however, in the proposed method cloud is free to choose any CS-recovery algorithm. For instance, through faster and simpler algorithms such as SL0, cloud can recover intermediate cipher three times faster than LP method \cite{17Mohimani2009}. Also, in \cite{18Wang_2013}, five keys were used which led to a further computational burden in ECG recorders.  In comparison with the very recent work, ``\textit{Kryptein}'', in which adaptive dictionary learning was used for generating sparsifying dictionary, in the current proposed work, any dictionary either fixed  or adaptive dictionaries can be used. By selecting fixed dictionaries, such as Wavelet transform family or discrete cosine transform (DCT), the heavy task of training dictionary can be removed \cite{19Xue_2017}.

\subsection{Experimental results}

\begin{figure}
	\centering
	\includegraphics[width=\linewidth]{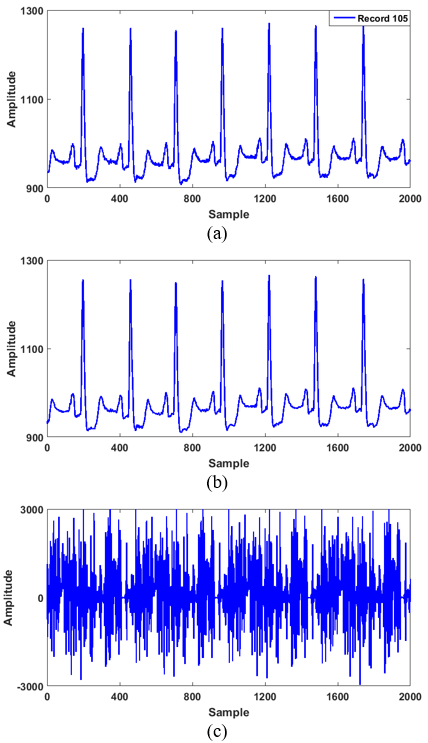}
	\caption{Recovery of ECG signal via DCT dictionary with and without actual key. (a) Initial ECG signal (plaintext).(b) Recovery of ECG Signal on user side with the actual key ($PRD= 0.29\%$, $SNR=50.6dB$). (c) Recovery of the ECG signal in cloud side with the wrong key ($PRD= 10^6\%$, $SNR= -30.22dB$).}
	\label{fig:fig01}
\end{figure}

The records from MIT-BIH Arrhythmia Database is used as ECG signal resource, \cite{47physiobank}. As ECG signal is used for diagnosis, the morphology and relative time positions of the various morphological features are important. We considered the diagnostic needs and proposed our method of encryption. Figure \ref{fig:fig01} shows the initial ECG signal (record number 105), recovered in cloud and then decrypted with and without the actual key. In this simulation, $2000$ samples of data record $105$ were chosen. Also, the deterministic binary block diagonal (DBBD) sensing matrix of size $\mathbf{A}_{128\times512}$ as suggested in \cite{DBBD} was used. DCT was used as sparsifying basis, and the first key $\mathbf{Q}$ was randomly chosen from a normal distribution $\mathcal{N}{(0,1/128)}$ to encrypt the measurement matrix. The second key $\mathbf{P}$ was randomly selected as a bipolar permutation key. According to the size of measurement matrix, the compression ratio is $N/M=4$. The important portion of diagnostic information of an ECG signal lies between its two consecutive QRS complexes, Fig. \ref{fig:fig01} shows that bipolar  permutation  conceals the diagnostic information, and without having the actual key, the decrypted signal is totally wrong.  
	
To evaluate the effectiveness of this approach on front of attacks, we simulated the process of estimating the bipolar permutation key. Four keys were chosen: one of them was the actual key and the others were $90\%$, $80\%$, and $70\%$ replica of actual key, i.e. $\mathbf{P}$, $\mathbf{E}^{90}$, $\mathbf{E}^{80}$ and $\mathbf{E}^{70}$. For instance, for the case of $90\%$, $90\%$ of the actual key’s components were copied into another matrix as estimated key and the remaining $10\%$ columns were randomly guessed. The simulation results verify that the bipolar permutation is strong enough for the application of ECG signal. The proposed approach was tested  with DCT and orthogonal Daubechies wavelets ($db 10$) dictionaries as these two are the fixed sparsifying dictionaries commonly used in the compressive sensing studies using ECG signals. In this simulation, $1024$ ECG samples of five different signals and a compression ratio, $CR=M/N=1/8$ were chosen. The results are shown in Table \ref{tab_4}. 

\begin{table}
	\renewcommand{\arraystretch}{1.5}
	\caption{Recovery by different keys and sparsifying basis(PRD\%) }
	\label{tab_4}
	\centering
	\begin{tabular}{|l|llll|llll|} 
		\hline
		\multirow{2}{*}{\rotatebox{90}{Records}} & \multicolumn{4}{l|}{DCT}                              & \multicolumn{4}{l|}{Wavelet}                         \\ 
		\cline{2-9}
		& $\mathbf{P}$ & $\mathbf{E}^{90}$ & $\mathbf{E}^{80}$ & $\mathbf{E}^{70}$ & $\mathbf{P}$ & $\mathbf{E}^{90} $ & $\mathbf{E}^{80}$ & $\mathbf{E}^{70}$  \\ 
		\hline
		100& 1.8&30.3&54.2&97.2&1.6&54.8&77.1&89.9\\
		101& 1.4&30.2&68.0&	105&	1.3&	54.0&	76.3&	93.6\\
		102& 1.2&2.1&16.8&	36.6&	1.2&	51.7&	73.6&	90.4\\
		103& 2.3&7.5&41.6&	61.0&	2.3&	52.7&	76.3&	90.2\\
		104& 1.3&16.8&58.7&	68.7&	1.3&	56.7&	76.6&	93.2\\
		105& 0.6&7.2&54.9&	84.0&	0.6&	55.3&	75.8&	91.6\\
		106& 2.6&1.7&41.1&	64.9&	1.8&	49.4&	72.2&	88.0\\
		107& 1.5&21.2&55.8&	95.3&	2.2&	53.3&	71.8&	87.6\\
		108& 0.4&15.4&59.3&	63.7&	0.4&	55.0&	73.6&	88.8\\
		109& 0.6&17.6&42.0&	84.5&	0.8&	52.1&	71.9&	86.0\\		
		\hline
	\end{tabular}
\end{table}
 
\begin{figure}
	\centering
	\includegraphics[width=\linewidth]{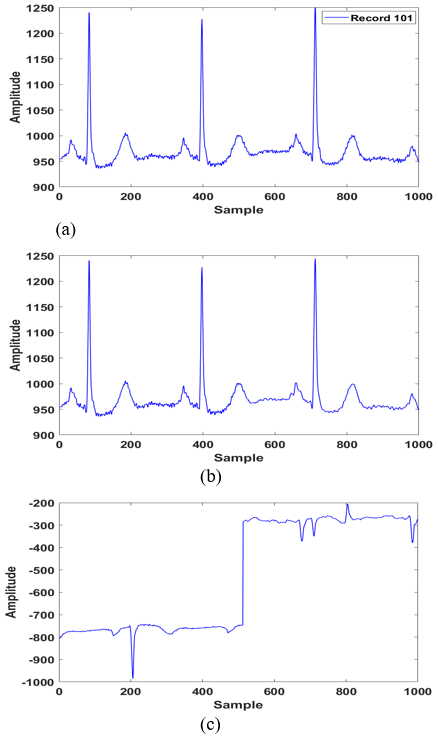}
	\caption{Recovery via adaptive dictionary learning (MOD) with and without actual key. (a) Initial ECG signal (plaintext). (b) Recovery of ECG Signal on user side with the true key $PRD = 0.09\%$, $SNR = 60.4dB$. (c) Recovery of the ECG signal in cloud side with the wrong key $PRD = 150\%$, $SNR= -3dB$. }
	\label{fig:fig04}
\end{figure}

In the aforementioned simulations, the fixed sparsifying dictionary was assumed to be available on the cloud side. On the other hand, if we employ adaptive dictionary learning, beside bipolar permutation matrix, sparsifying dictionary will also be unknown for the "curious" cloud or attacker. Hence, if  adaptive dictionary learning were used, the secrecy of the system can be increased. Adaptive dictionaries usually yield higher quality in reconstruction at the expense of computational burden to the system. Since, the learning process needs to be executed only once for a subject, this complexity may be conveniently ignored. There are numerous adaptive dictionary learning methods such as method of optimal direction MOD \cite{48Engan2000}, and K-singular value decomposition (K-SVD) \cite{49Aharon2006}.  In order to demonstrate the proposed method with the adaptive dictionary learning, MOD, which is one of the fastest method to learn sparsifying dictionary was chosen. Figure \ref{fig:fig04} shows the result obtained using the adaptive sparsifying dictionary while using the ECG signal (record number 101) from the MIT-BIH Arrhythmia database. It is evident that recovery without key, or recovery of encrypted signal leads to totally wrong recovery where no information of the original ECG signal can be seen. 
 
Proposed method does not affect the quality of the reconstruction.  In section \ref{sec_Proposed_method} it was shown that after recovery in cloud, end user should be able to exactly recover the initial signal at their end. This aspect of the proposed  method was also tested for a number of ECG signals from the database, (records no. $100-104$). Table \ref{tab_5} shows that there is no difference in the quality of the reconstructed signal with and without proposed encryption system.

\begin{table}
	\renewcommand{\arraystretch}{1.5}
	\caption{Impact on the quality of reconstruction }
	\label{tab_5}
	\centering
	\begin{tabular}{|l|ll|ll|ll|} 
		\hline
		\multirow{3}{*}{\rotatebox{90}{Records}} & \multicolumn{6}{c|}{SNR(dB)}\\ 
		\cline{2-7}
		& \multicolumn{2}{c|}{CR=50\%} & \multicolumn{2}{c|}{CR=75\%} & \multicolumn{2}{c|}{CR=87.5\%}  \\
		& Ordinary& Secure & Ordinary & Secure & Ordinary&Secure\\ 
		\hline
		100&	57.02&	57.02&	45.56&	45.56&	35.28&	35.28\\
		101&	58.35&	58.35&	47.14&	47.14&	35.81&	35.81\\
		102&	56.21&	56.21&	44.16&	44.16&	38.24&	38.24\\
		103&	60.08&	60.08&	49.44&	49.44&	33.23&	33.23\\
		104&	52.15&	52.15&	42.44&	42.44&	37.31&	37.31\\
		\hline
	\end{tabular}
\end{table}

\begin{figure}[!ht]
    \centering
    \includegraphics[height = 2.2in, width = 3.4in]{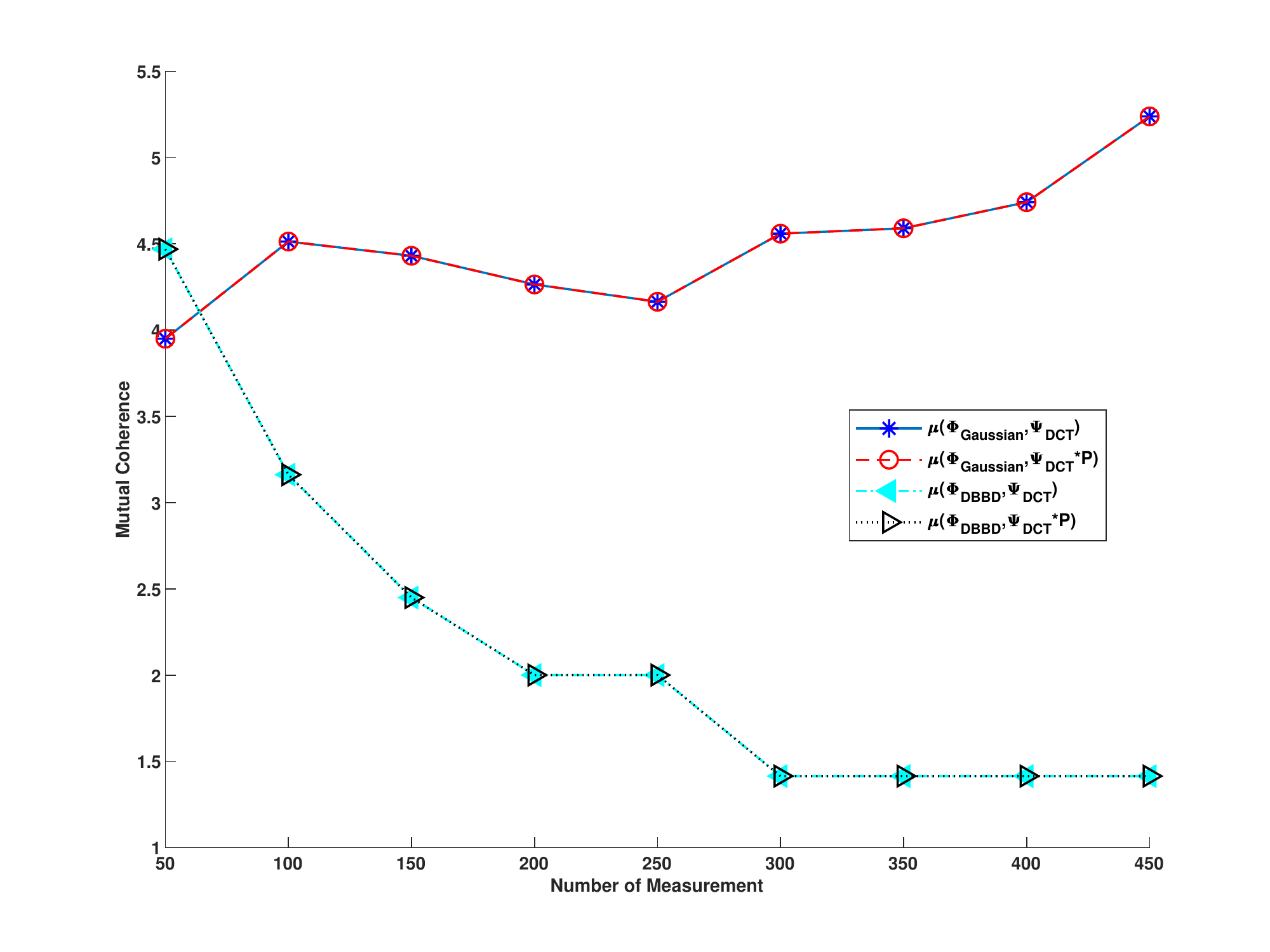}
    \caption{Impact of proposed encryption on mutual coherence.}
    \label{fig:fig05}
\end{figure}

Also, any change in mutual coherence can be reflected to the quality of reconstruction \cite{09Zanddizari2018,DBBD,50Yan2014}. With this regard, we checked the effect of proposed method on the mutual coherence. Figure \ref{fig:fig05} shows this effect for random and deterministic measurement matrices as a function of number of measurements. For the class of random measurement matrices, we generated by a zero-mean and variance $1/M$ i.i.d. Gaussian process, denoted by $\Phi_{Gaussian}$. For the class of deterministic measurement matrices, DBBD matrix $\Phi_{DBBD}$ was used.  DCT matrix $\Psi_{DCT}$ and encrypted DCT matrix $\Psi_{DCT}*\mathbf{P}$ were used as sparsifying and encrypted sparsifying dictionary, respectively. In this simulation, $N=500$ and the number of measurements was changed to check the effect of mutual coherence on different sizes of matrices. The results show the bipolar permutation matrix does not affect the mutual coherence.

To the best of knowledge of the authors, there has been no specific work on secure CS recovery outsourcing for the ECG signal. However, there are two methods in the literature, namely OIRS and Kryptein, that are related to the proposed method for secure CS recovery outsourcing. The proposed work was compared with these related methods under the following considerations:  recovery algorithms, sparsifying bases and  computational complexity.  Table \ref{tab_6} shows comparison of the proposed method with the ORIS and Kryptein.  The proposed method can be applied for any CS recovery algorithm and has low overload both on the user side and the cloud.  

\begin{table}[!ht]
	\renewcommand{\arraystretch}{1.5}
	\caption{Comparison of functionality}
	\label{tab_6}
	\centering
	\begin{tabular}{p{30mm}|ccc}
	\hline
	Functionality & OIRS \cite{18Wang_2013}& Kryptein\cite{19Xue_2017} & Proposed\\\hline  
     Recovery based on LP methods   
	 & Yes  & Yes & Yes	   \\
	 Recovery based on Matching pursuit, Belief Propagation, and Sl0
	 & No  & Yes & Yes	   \\
	 Using DCT/Wavelet sparsifying dictionaries	
	 & Yes	& No	& Yes \\
	 Using adaptive sparsifying dictionary	
	 & Yes	& Yes	& Yes \\

	 Complexity in User end (multiplication operation)	
	 &$4N^2$	&$N^2$	&$N^2$ \\
		\hline
	\end{tabular}
\end{table}

\subsection{Complexity of the proposed method}

The proposed method can be categorized  as  a  fast  and  energy  efficient method of encryption. Proposed method requires two keys; first key is a random square matrix and the second  key  is  a bipolar permutation  matrix.  The  first  key  is  used  to encrypt  the  measurement  matrix.  As sensors  might  use  deterministic  or  structural measurement matrices in certain applications,  attacker may use the structure in measurement matrix and consequently detect the bipolar permutation matrix. When using deterministic measurement matrices for the recovery service, cloud has  $\mathbf{A}^*$ ,  where $\mathbf{A}^*= \hat{\mathbf{A}} \mathbf{P} = \mathbf{Q} \mathbf{A} \mathbf{P}=\mathbf{Q} \Phi\Psi\mathbf{P}$. Without the first key, for the case where $\Phi$ is deterministic, attacker can separate $\Psi \bP$ from $\Phi\Psi\bP$ and since $\Psi$ are known-such as DCT or wavelet dictionary- the permutation matrix may be revealed. But, if the first key is applied in addition, this attack can be avoided. Suppose the first key to be chosen from Gaussian distribution as it has maximum entropy that causes maximum diffusion. The overload of the first key is just $M \times M$ multiplications and $M \times (M-1)$ addition operations for sending each measurement vector.  The measurement matrix that is shared with the cloud would be  $\mathbf{Q}\Phi\Psi\mathbf{P}$  instead of $\mathbf{Q}\Phi\Psi$. It leads to $N$ random shift in the columns of $\mathbf{Q}\Phi\Psi$ and its components are randomly multiplied by $-\alpha$ or $+\alpha$. The matrix $\mathbf{Q}\Phi\Psi\mathbf{P}$ must be available in cloud to do the recovery process. {To further enhance privacy, every individual user would have a unique key. Also, after certain number of queries, to prevent potential known plaintext attack (KPA), the key can be updated.} 

\section{Conclusion}
\label{sec_conclusion}

For doing CS-recovery service in cloud environment, the secrecy of information should be preserved. When ECG measurements are transmitted to the cloud, the cloud with its strong resources can do the CS recovery for the client. Through the proposed method, not only does the cloud conduct the CS-recovery, but post recovery it also delivers an encrypted version of signal, thereby preserving the privacy of patient’s information during the entire process. The proposed encryption is carried out in the sparse domain. Through a bipolar permutation matrix, the initial sparse vector (plaintext) is mapped into to another sparse vector (cipher). The cloud after recovery presents a permuted sparse vector to the user and without knowing the key, it would be very difficult to guess the original signal as the degree of freedom for this guess is small. In other words, with respect to the ECG signal where small changes might distort the signal, it is practically difficult to guess the information contained in the signal for "curious" cloud or eavesdropper. The role of the sparsifying basis in improving the secrecy of information is also demonstrated in this study. Appropriate choice of adaptive sparsifying basis can also provide additional secrecy.

\bibliographystyle{IEEEtran}



\begin{IEEEbiography}[{\includegraphics[width=1.1in,height=1.1in,clip,keepaspectratio]{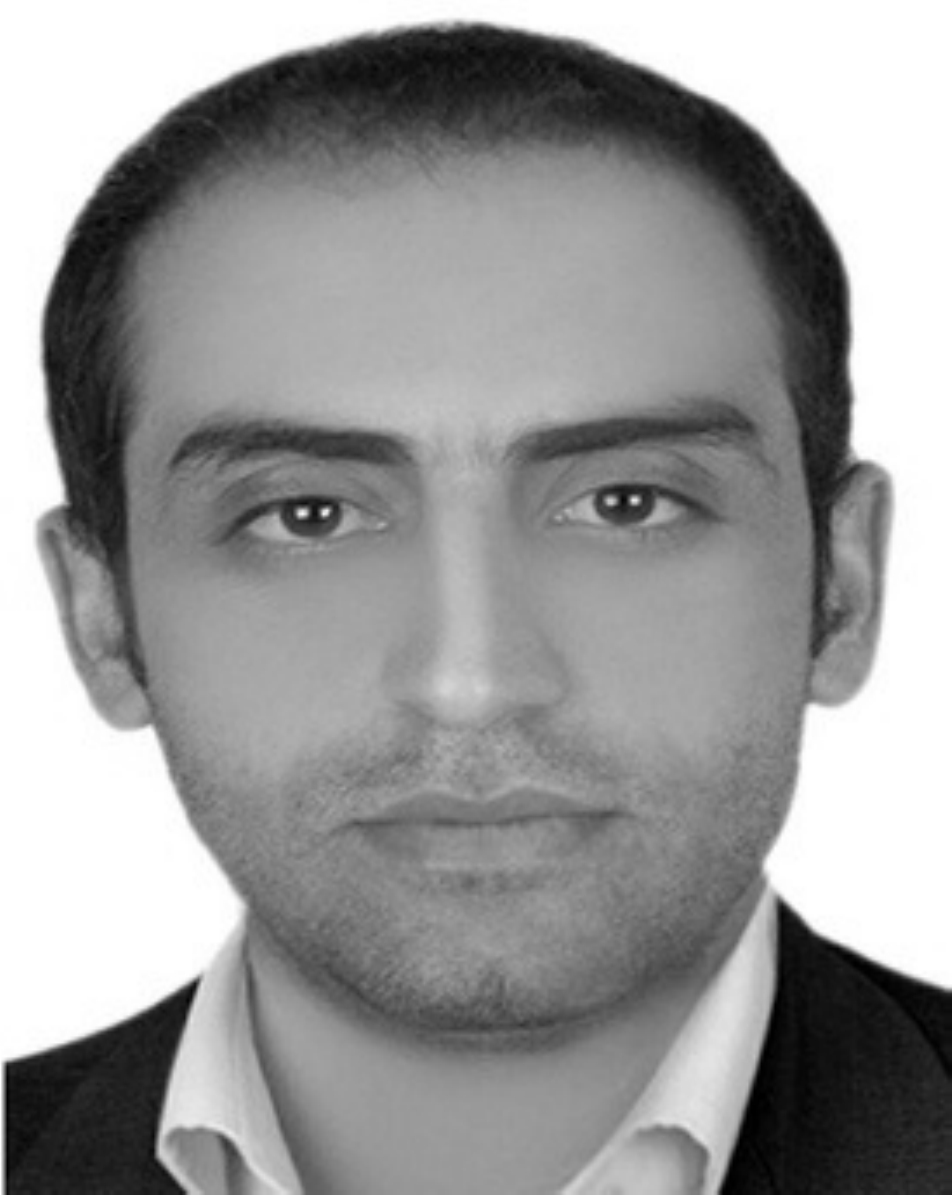}}]{Hadi Zanddizari} (S'19) received Master’s in Electrical Engineering from University of Science and Technology, Iran, in 2015. From 2015 to 2017 he worked at Pardis Tehnology Park as a researcher and
programmer. Currently, he is working towards the PhD degree from the University of South Florida. His research focuses are on sparsity, compressive sensing, cybersecurity, machine learning, and robustness of machine learning algorithms.
\end{IEEEbiography}

\begin{IEEEbiography}[{\includegraphics[width=1.5in,height=1.25in,clip,keepaspectratio]{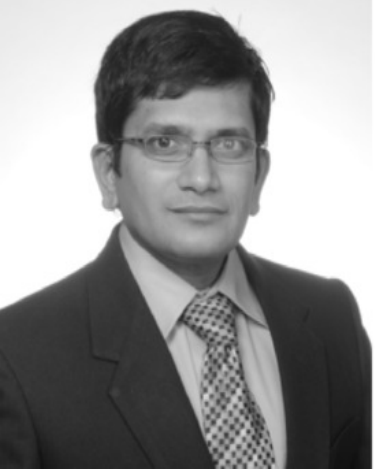}}]{Sreeraman Rajan} (M'90-SM'06) is a Canada Research Chair in Sensor Systems in the Department of Systems and Computer Engineering in Carleton University, Ottawa, Canada since 2015. He is also currently the Associate Director, Ottawa Carleton Institute for Biomedical Engineering.  Before joining Carleton University, he was with Defence Research and Development Canada (DRDC) Ottawa, Canada as a Senior Defence Scientist. His  industrial experience includes areas of nuclear science and engineering, control, electronic warfare, communications and biomedical engineering in addition to research experience in areas of sensor signal/image processing, pattern recognition and machine learning.   He is currently the Chair of the IEEE Ottawa EMBS and AESS Chapters.  He has served IEEE Canada as its board member (2010-Oct 2018) and the IEEE MGA in its Admissions and Advancement Committee, Strategic and Environment Assessment Committee.  He was awarded the IEEE MGA Achievement Award in 2012 and recognized for his IEEE contributions with Queen Elizabeth II Diamond Jubilee Medal in 2012. IEEE Canada recognized his outstanding service through 2016 W.S.  Read Outstanding Service Award.  IEEE Ottawa Section recognized him as an Outstanding Volunteer in 2012 and an Outstanding Engineer in 2018.  He has been involved in organizing several successful IEEE conferences and has been a reviewer for several IEEE journals and conferences.  He is the holder of two patents and two disclosures of invention. He has authored more than 150 journal articles and conference papers. He is a Senior Member of IEEE, member of IEEE Instrumentation and Measurement, Engineering in Medicine and Biology, Signal Processing and Aerospace and Electronic Systems Societies.
\end{IEEEbiography}

\begin{IEEEbiography}[{\includegraphics[width=1.1in,height=1.1in,clip,keepaspectratio]{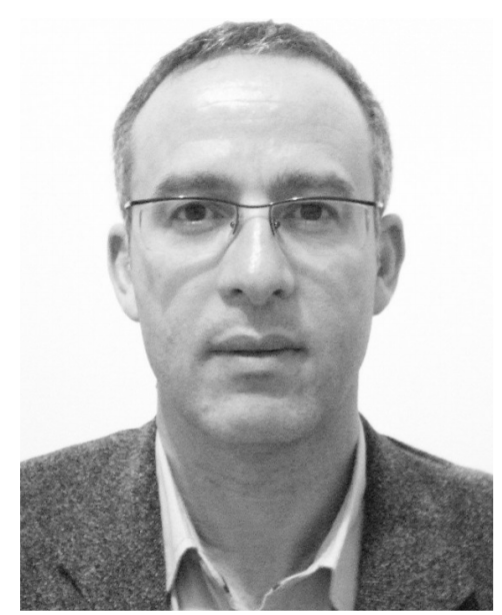}}]{Hassan Rabah} (M’06–SM’13) received the M.S. degree in electronics and control engineering and the Ph.D. degree in electronics from Henri Poincaré University, Nancy, France, in 1987 and 1993, respectively. He became an Associate Professor of Electronics Microelectronics and Reconfigurable Computing with the University of Lorraine, Nancy, in 1993,and a Full Professor in 2011. In 1997, he joined the Architecture Group of LIEN, where he supervised research on very large-scale integration implementation of parallel architecture for image and video processing. He also supervised research on the field-programmable gate array (FPGA) implementation of adaptive architectures for smart sensors in collaboration with industrial partners. He participated in several national projects for quality of service measurement and video transcoding techniques. In 2013, he joined the Institut Jean Lamour, University of Lorraine, where he is currently the Head of the Measurement and Electronics Architectures Group. His current research interests include partial and dynamic reconfigurable architectures for adaptive systems, design, implementation of FPGA-based embedded systems with an emphasize on power optimization, video compression decompression and transcoding, compressive sensing, and sensor networks.Dr. Rabah has been a Program Committee Member for a number of conferences.
\end{IEEEbiography}

\begin{IEEEbiography}[{\includegraphics[width=1.1in,height=1.1in,clip,keepaspectratio]{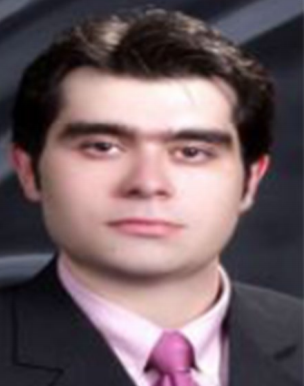}}]{Houman Zarrabi} received his doctoral of engineering from Concordia University in Montreal, Canada in 2011. Since then he has been involved in various industrial and research projects. His main expertise includes IoT, M2M, big data, embedded systems and VLSI. He is currently the national IoT program director and a faculty member of ITRC.
\end{IEEEbiography}

\vfill

\end{document}